\documentclass[aps,prl,twocolumn,superscriptaddress]{revtex4-2}
\usepackage{amsmath}
\usepackage{mathtools}
\usepackage{amssymb}
\usepackage{mathrsfs}
\usepackage{dsfont}
\usepackage[pdftex,colorlinks=true,allcolors=blue,breaklinks=true,bookmarks=false]{hyperref}
\usepackage{graphicx}
\usepackage{float}
\usepackage{xcolor}

\usepackage{multibib} %right-align reference numbers [for arXiv]

    \usepackage{titlesec}
    \titleformat*{\section}{\large
    \bfseries}
    \titleformat*{\subsection}{\bfseries}
    \titleformat*{\subsubsection}{\itshape}

\begin{document}

\title{Frustrated Quantum Magnetism on Complex Networks: What Sets the Total Spin}

\author{Preethi Gopalakrishnan}
%\email[E-mail: ]{preethi.gopalakrishnan@uni.lu}
%\thanks{Present address: Department of Physics and Materials Science, University of Luxembourg, L-1511 Luxembourg.}
\affiliation{School of Physics, IISER Thiruvananthapuram, Kerala 695551, India}
\author{Shovan Dutta}
%\email[E-mail: ]{shovan.dutta@rri.res.in}
\affiliation{Raman Research Institute, Bangalore 560080, India}

%\date{\today}

\begin{abstract}
Consider equal antiferromagnetic Heisenberg interactions between qubits forming a complex, nonbipartite network. We ask the question: How does the network topology determine the net magnetization of the ground state and to what extent is it tunable? By examining over $75000$ networks of different families with tunable structural properties, we demonstrate that (i) heterogeneity in the number of neighbors is essential for a nonzero total spin, and (ii) apart from the number of neighbors, the key determinant is the presence of (disassortative) hubs, as opposed to the frustration level. In fact, one can vary the magnetization throughout its range by embedding such hubs. We also discuss simple, exactly solvable networks where such tunability leads to both abrupt and continuous transitions, with quantum effects giving rise to a diverging susceptibility. Our findings can be realized on emerging platforms and pose a number of fundamental questions, strongly motivating wider exploration of quantum many-body phenomena on complex networks.
\end{abstract}

\maketitle

Understanding how a network of interactions between dynamical units control their collective behavior is a unifying challenge for modern science. While, traditionally, condensed matter physics has focused on short-range interactions on a lattice, the rapidly growing field of complex networks \cite{Strogatz2001, Newman_book, MiniReview2020} has shown that the network topology itself can play a pivotal role \cite{Dynamics_book}. In particular, structures absent in regular lattices, e.g., heterogeneity in the number of neighbors, the small-world effect \cite{WattsStrogatz1998}, dissimilarity between adjacent nodes (called ``dis-assortativity'' \cite{Newman2002}), and communities \cite{Newman2011}, can radically alter critical phenomena, disease spreading, synchronization, response to node failures, and other dynamical processes \cite{Dynamics_book, 2008_RMP, 2006_PhysRep, 2002_RMP_BA, Explosive_2019, Explosive_2016, Kuramoto_2016}. %Furthermore, interplay between topological and dynamical variables can yield new phenomena such as explosive synchronization and other abrupt transitions \cite{Explosive_2019, Explosive_2016, Kuramoto_2016}. 

However, most of these findings are for classical systems. While there is a sizable literature on single-particle quantum dynamics \cite{CTQW_2011, Localization_2008, RRG_2021, QuantumNet_2019, Quantum_2023} and quantum communication networks \cite{QuantumNet_2019, Quantum_2023}, the exploration of quantum \emph{many-body} networks is in its infancy \cite{Quantum_2023, BH_2012, JCH_2013, NoSW_2020, Daley_2019, SW_Scrambling_2019, Chaos_Ising_Adolfo_2025, Sundar2021, Entanglement_Ising_2025}. In particular, there is no general understanding of how network structure governs collective properties, especially of a frustrated quantum system. This question is timely, as it would soon be possible to design essentially arbitrary interaction graphs between qubits with long coherence times, most notably with trapped ions \cite{Monroe2021, Korenblit2012, Teoh2020, Davoudi2020} and superconducting circuits \cite{Lamata2018, Tsomokos2010, Kollar2019, Onodera2020, King2023}, among other setups \cite{Zoller_2015, Zoller_2020, Hung2016, McMahon2016, Fung2023, Schlipf2017}.

In the context of magnetism, networks can encode variable levels of geometric frustration, which is at the heart of exotic phases of matter such as spin liquids \cite{Intro_FM_book, Diep_book}. Frustration arises whenever all bonds cannot be simultaneously satisfied, the simplest example being a triangle with antiferromagnetic bonds \cite{Moessner2006}. Studies of frustration in network science have focused on social imbalance \cite{SocialBalance_2014} and Ising spin-glass physics \cite{2008_RMP}. In contrast, here we consider an antiferromagnetic Heisenberg model of qubits, $\hat{H} = \sum_{\langle i,j \rangle} J_{i,j} \hat{\mathbf{S}}_i \cdot \hat{\mathbf{S}}_j$, where we generally take $J_{i,j} = 1$ to focus on the role of network topology as opposed to bond disorder. For a bipartite graph, the two sublattices (of size $N_A$ and $N_B$) align oppositely, yielding a total spin $S_{\text{total}} = |N_A - N_B|/2$ \cite{Lieb1962}. On the other hand, for a nonbipartite lattice (e.g., Kagome) frustration can stabilize a quantum spin liquid composed of fluctuating singlets with $S_{\text{total}} = 0$ but no conventional long-range order~\cite{QSL_2017, QSL_RMP_2017, QSL_2023}. %Very little is known about the ordering of complex nonbipartite graphs. 
Here we explore what sets $S_{\text{total}}$ in a complex nonbipartite graph and to what extent it is tunable. Although far from fully characterizing the ground state, our findings reveal surprising ways in which network topology shapes collective behavior of such systems.

We examine more than $70$ network families of $30$ qubits with adjustable structural properties, each comprising an ensemble of $1000$ graphs, using state-of-the-art exact diagonalization based on the \texttt{HPhi} library \cite{HPhi1, HPhi2}. We find, surprisingly, that $S_{\rm total}$ is not sensitive to the level of frustration; however, it is fully tunable by embedding ``hubs’’ connected to many low-degree nodes, allowing one to control spin states by varying the graph topology. We also discuss exactly solvable cases where quantum effects enhance the sensitivity to network parameters, producing a divergent susceptibility. Our results show that the role of heterogeneity in promoting spin alignment \cite{Dynamics_book} extends to quantum systems, while uncovering the central importance of assortative correlations. They also constrain the magnetic ordering on complex networks and raise fundamental questions on the wider role of frustration.

\begin{figure}
    \centering
    \includegraphics[width=1\columnwidth]{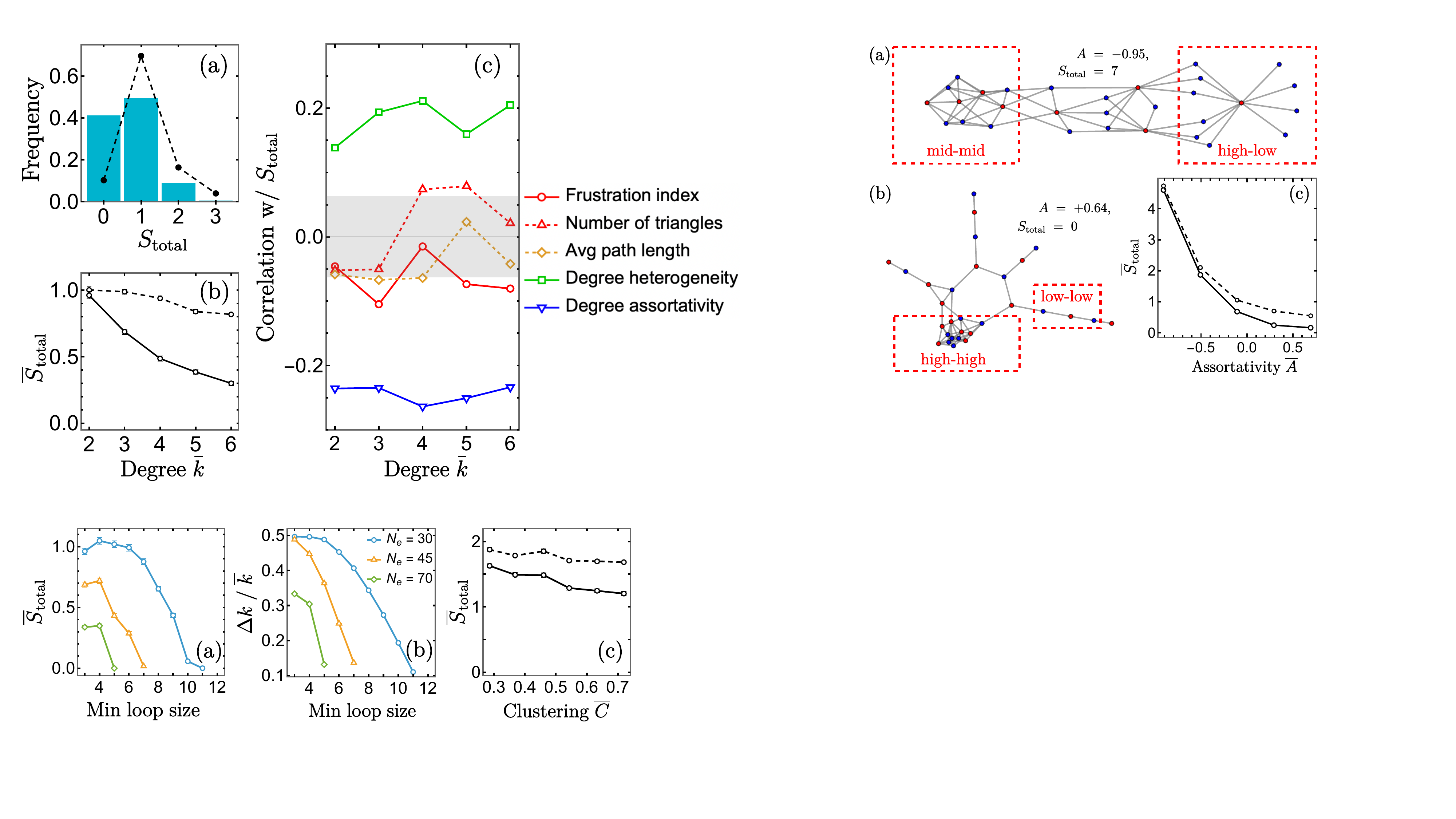}
    \caption{(a) Distribution of total spin in the ground state of random graphs with $N = 30$ qubits and $N_e = 45$ bonds with equal antiferromagnetic Heisenberg coupling. Dashed curve shows the average magnetization, $S^z_{\text{total}}$, for Ising spins (max-cut bipartitions). (b) Ensemble average of $S_{\text{total}}$ vs average degree $\bar{k} = 2 N_e / N$. (c) Correlation with various graph metrics. Points inside the gray band are statistically insignificant.
    }
    \label{fig:randgraph}
\end{figure}

\textit{Random graphs}|We start by examining (connected) Erd\H{o}s--R\'enyi graphs of \(N\) sites and \(N_e\) bonds. Figure~\ref{fig:randgraph}(a) shows the distribution of such an ensemble. We notice that \(S_{\text{total}}\) is small compared to its upper bound, \(N/2\), which one might expect as all bonds have antiferromagnetic coupling. Second, as shown in Fig.~\ref{fig:randgraph}(b), the average magnetization $\bar{S}_{\text{total}}$ falls with $N_e$ or, equivalently, the average degree \(\bar{k} = \frac{2N_e}{N}\). This trend is consistent across all network families and interpolates between known limiting cases: For \(N_e^{\text{min}} = N-1\) one gets a random tree where a sublattice imbalance gives rise to a net magnetization with $\bar{S}_{\text{total}} \approx 0.23 \sqrt{N}$; see Supplemental Material (SM) \cite{SuppMat}. On the other hand, \(N_{e}^{\text{max}}=N(N-1)/2\) gives an all-to-all graph, where \(\hat{H} = \frac{1}{2}\hat{\mathbf{S}}_{\text{total}}^2 \) up to a constant and the ground state has \(S_{\text{total}}=0\).

More importantly, Fig.~\ref{fig:randgraph}(c) shows that $S_{\text{total}}$ is not sensitive to the frustration level of a graph but rather to its heterogeneity and assortativity \cite{Newman2002}. We look at three measures of frustration \cite{Aref2017}: (1) the frustration index, which is the minimum number of bonds one needs to cut to make the graph bipartite  \cite{Bipartivity_2003, Aref2017, Aref2018}, (2) the number of triangles, which is a more local measure, and (3) the average path length (number of hops) between two nodes, which controls how strongly a given spin can affect any other spin in the network, adding to the level of frustration. Surprisingly, we find that all three measures have weak or no correlation with $S_{\text{total}}$. Instead, $S_{\text{total}}$ is correlated with the amount of heterogeneity and assortativity: The former is quantified by $\Delta k$, the spread in the number of neighbors \cite{Dynamics_book}, whereas the latter keeps track of degree-degree correlation, i.e., whether high-degree nodes connect to high-degree nodes $(A>0)$ or to low-degree nodes $(A<0)$ \cite{Newman2002}. Figure~\ref{fig:randgraph}(c) tells us that $S_{\text{total}}$ is larger for more heterogeneous graphs where high-degree nodes link to low-degree nodes. Below we vary these structural properties in turn to make sense of the correlations.

We also find a strong correlation ($\sim 0.6$) with the magnetization, $S^z_{\rm total}$, of Ising spins on the same graph. The latter is given by an optimal bipartition (known as the max-cut problem \cite{2008_RMP}), where one divides the spins into oppositely aligned groups $A$ and $B$ so as to maximize the number of $A$-$B$ bonds. As shown in Fig.~\ref{fig:randgraph}, such a bipolar state overestimates $S_{\text{total}}$ for the Heisenberg spins, especially at larger $\bar{k}$. Nonetheless, the major variations with structural properties are reflected in this purely topological metric. Note the solution to the max-cut problem is generally not unique, in which case we take the average $S^z_{\rm total}$ over the degenerate manifold.

\begin{figure}
    \centering
    \includegraphics[width=1\columnwidth]{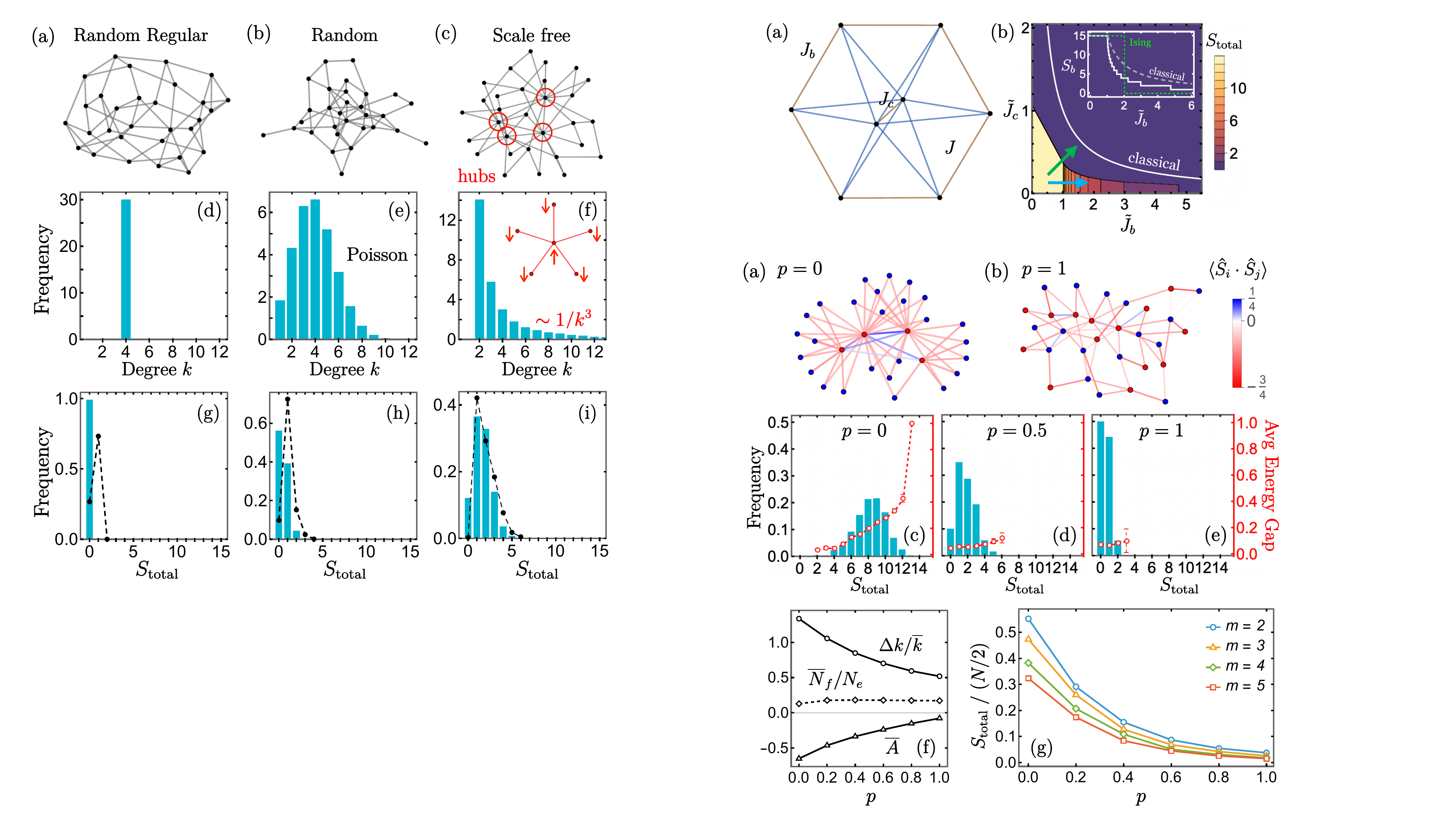}
    \caption{Sample graph (top), degree distribution (middle), and $S_{\text{total}}$ distribution (bottom) for random regular, random, and scale-free (Barab{\'a}si-Albert) graphs with $N = 30$ and $\bar{k} \approx 4$. The power-law tail in (f) gives rise to hubs [circled in (c)] that want to polarize their neighborhoods (inset). Dashed curves in (g-i) show the distribution of average $S^z_{\rm total}$ for Ising spins.
    }
    \label{fig:heterogeneity}
\end{figure}

\textit{Importance of heterogeneity}|Suppressing heterogeneity results in the family of random regular graphs, where each spin has the same degree $k$. Figure~\ref{fig:heterogeneity}(g) shows that for $k = 4$ one almost always has $S_{\text{total}} = 0$, and the same holds for $k \neq 4$ (see SM \cite{SuppMat}). This strong result shows that heterogeneity is essential for a nonzero total spin. 

Conversely, one can ramp up the heterogeneity by generating scale-free graphs \cite{SF_book} where the degree distribution $p_k$ follows a power law. A well known example is the Barab{\'a}si-Albert model \cite{BarabasiAlbert1999} for which $p_k \sim 1/k^3$. As shown in Fig.~\ref{fig:heterogeneity}(i), these graphs exhibit much higher magnetizations. The enhancement can be understood as originating from ``hubs,'' i.e., nodes with a disproportionately high degree. Such hubs form a locally star structure, favoring a net spin imbalance [see Fig.~\ref{fig:heterogeneity}(f) inset]. As we discuss below, this effect persists even when hubs are embedded in a broader class of networks.

Note that all three families in Fig.~\ref{fig:heterogeneity} have similar frustration indices ($\sim 10.1$, $10.5$, and $10.6$, respectively).

\begin{figure}
    \centering
    \includegraphics[width=1\columnwidth]{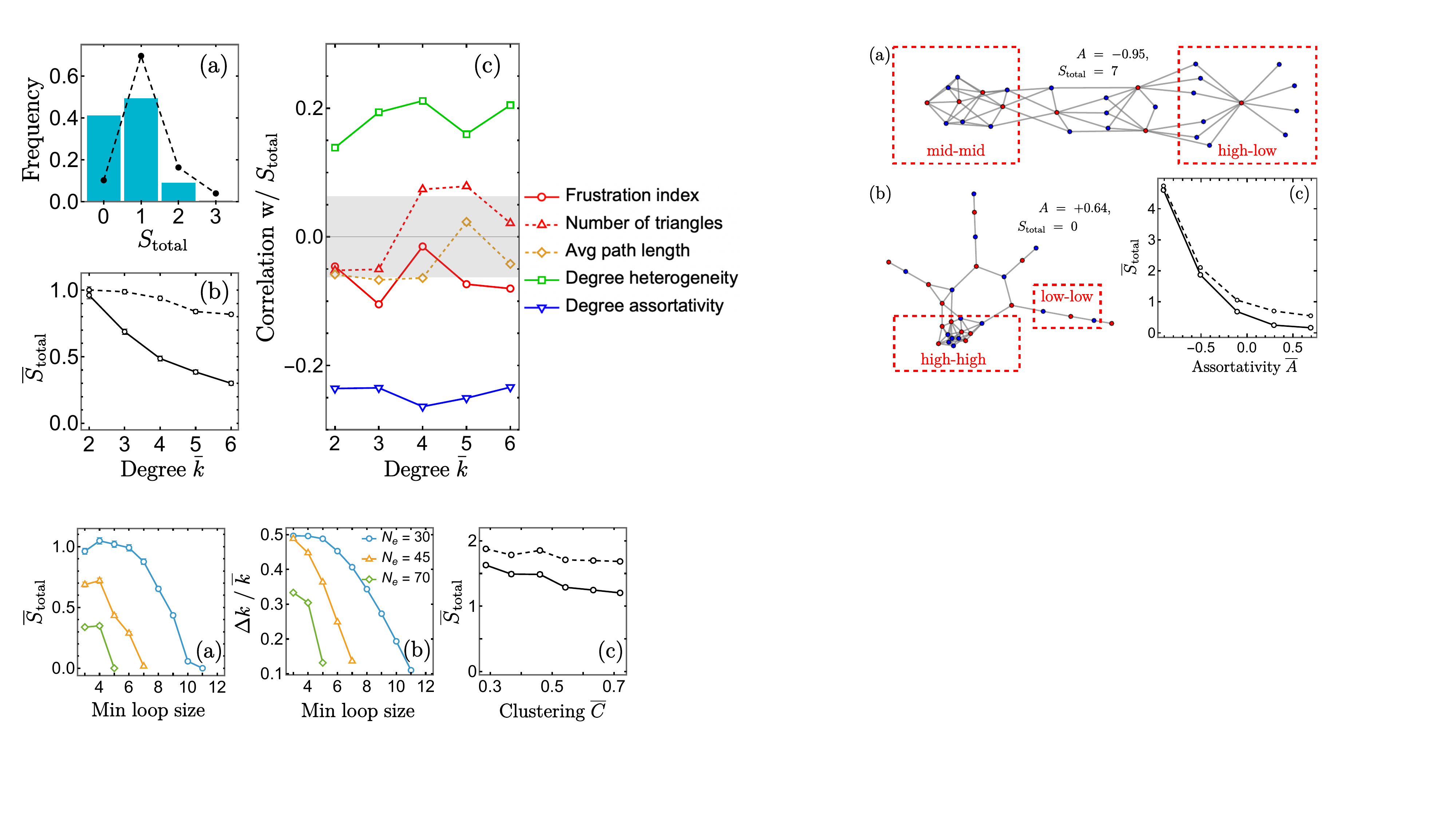}
    \caption{(a) Average total spin and (b) heterogeneity, given by the standard deviation $\Delta k$ of the degree distribution, for random graphs without short loops and $N = 30$. (c) $\bar{S}_{\text{total}}$ for Barab{\'a}si-Albert graphs with $N = 30$, $\bar{k} = 3.8$, and tunable number of triangles, measured by the mean clustering coefficient $C \in [0,1]$ \cite{WattsStrogatz1998}. Dashed curve shows $\bar{S}^z_{\text{total}}$ for Ising spins.
    }
    \label{fig:frustration}
\end{figure}

\textit{Insensitivity to frustration}|An alternative argument for the unmagnetized nature of random regular graphs [Fig.~\ref{fig:heterogeneity}(g)] could be that they have fewer short loops compared to random graphs \cite{Wormald1981, Bonneau2017}, so they appear locally treelike. Thus, $S_{\text{total}}$ is small due to lack of local frustration, not homogeneity. To test this proposition, we vary the number of short loops in two different ways: First, we uniformly sample random graphs with a minimum loop size (girth) $l$ \cite{Bayati2018}. We find that the total-spin distribution is almost unaffected by removing all triangles. As $l$ is increased, $\bar{S}_{\text{total}}$ eventually drops to zero close to the maximum girth $l_{m}$ [Fig.~\ref{fig:frustration}(a)]; however, $l_{m}$ depends strongly on the number of bonds $N_e$ \cite{Verstrate2016}, and $\bar{S}_{\text{total}}$ does not appear to be sensitive to whether $l$ is even or odd. Instead, it follows the variation in the degree distribution, which becomes progressively narrower with the removal of short loops [Fig.~\ref{fig:frustration}(b)]. Thus, we infer that $S_{\text{total}}$ is dictated by the heterogeneity and not by the frustration. Second, we follow Ref.~\cite{Holme2002} to augment the Barab{\'a}si-Albert model with a triangle formation step, which allows one to tune the number of triangles without affecting the degree distribution or assortativity. As shown in Fig.~\ref{fig:frustration}(c), this yields a weak variation of $\bar{S}_{\text{total}}$. %which is reproduced by the max-cut bipartitions. 
We also find that $\bar{S}_{\text{total}}$ does not vary significantly with shorter path lengths in small-world networks (see End Matter).

While this indifference to frustration may be counterintuitive, note the same holds on regular lattices, where both non-frustrated (square) and frustrated (Kagome) cases have $S_{\text{total}} = 0$, albeit with distinct spin ordering. %Likewise, we expect frustration to be important in setting the order on general networks. 
However, its extension to general networks is indeed surprising and demands a deeper explanation.

\begin{figure}
    \centering
    \includegraphics[width=1\columnwidth]{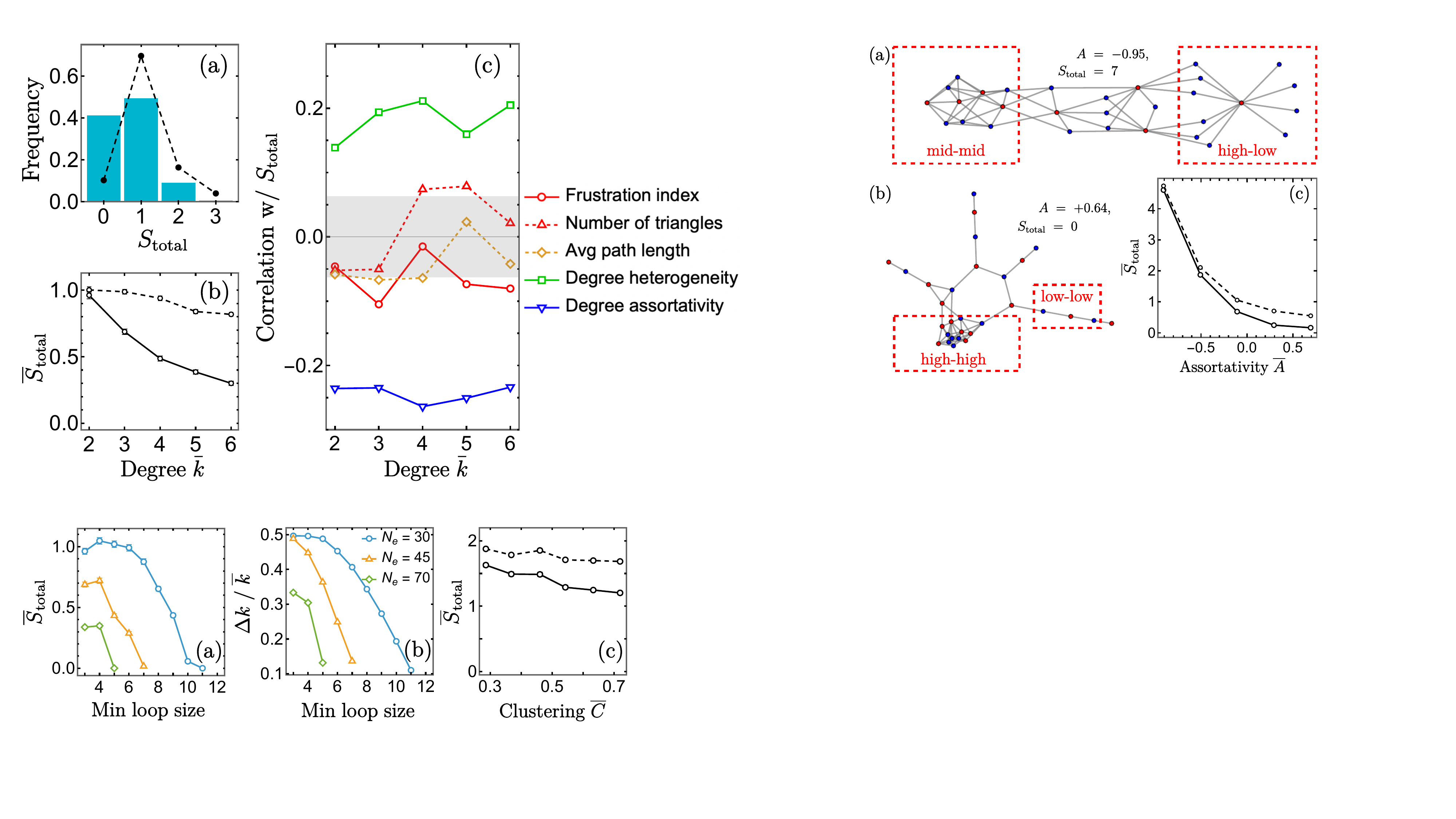}
    \caption{(a) Most disassortative and (b) most assortative networks with the same set of degrees. Node colors show a bipolar ground state of Ising spins. (b) Variation of $\bar{S}_{\text{total}}$ as the assortativity \cite{Newman2002} is tuned for random graphs with $N = 30$ and $\bar{k} = 4$. Dashed curve shows $\bar{S}^z_{\rm total}$ for Ising spins.
    }
    \label{fig:assortativity}
\end{figure}

\textit{Importance of disassortativity}|Not all heterogeneous networks have large magnetization. Another key metric is the assortativity coefficient $A \in [-1,1]$ \cite{Newman2002}. Perfect assortativity requires all nodes to have the same degree (for which $S_{\text{total}} \simeq 0$), whereas perfect disassortativity occurs in complete bipartite graphs with sublattice imbalance \cite{VanMieghem2010} (where $S_{\text{total}}$ can be large). Crucially, one can tune $A$ without altering the degrees via rewiring pairs of bonds \cite{VanMieghem2010, Sokolov2005, Noldus2015}. Thus, Figs.~\ref{fig:assortativity}(a) and \ref{fig:assortativity}(b) show two graphs with very different values of $A$ but the same degrees (hence, the same heterogeneity). The first graph is the most disassortative and has $S_{\text{total}} = 7$, whereas the latter is the most assortative and has $S_{\text{total}} = 0$. The physical origin of this large disparity becomes clear if one examines the structure of these networks \cite{Sokolov2005}: Strongly disassortative graphs have a section where few high-degree nodes (hubs) connect to many low-degree nodes in an approximately bipartite structure [Fig.~\ref{fig:assortativity}(a)], which gives rise to a large spin imbalance. By contrast, in strongly assortative graphs high-degree nodes form a tightly connected group and low-degree nodes form a treelike structure [Fig.~\ref{fig:assortativity}(b)], neither of which favors a large magnetization. Note that the large $S_{\text{total}}$ in the former case requires both the presence of hubs and disassortativity.

Figure~\ref{fig:assortativity}(c) shows that $\bar{S}_{\text{total}}$ falls sharply as we tune $A$ from its minimum to maximum value for random graphs. The same is observed for scale-free graphs (End Matter).

\begin{figure}
    \centering
    \includegraphics[width=1\columnwidth]{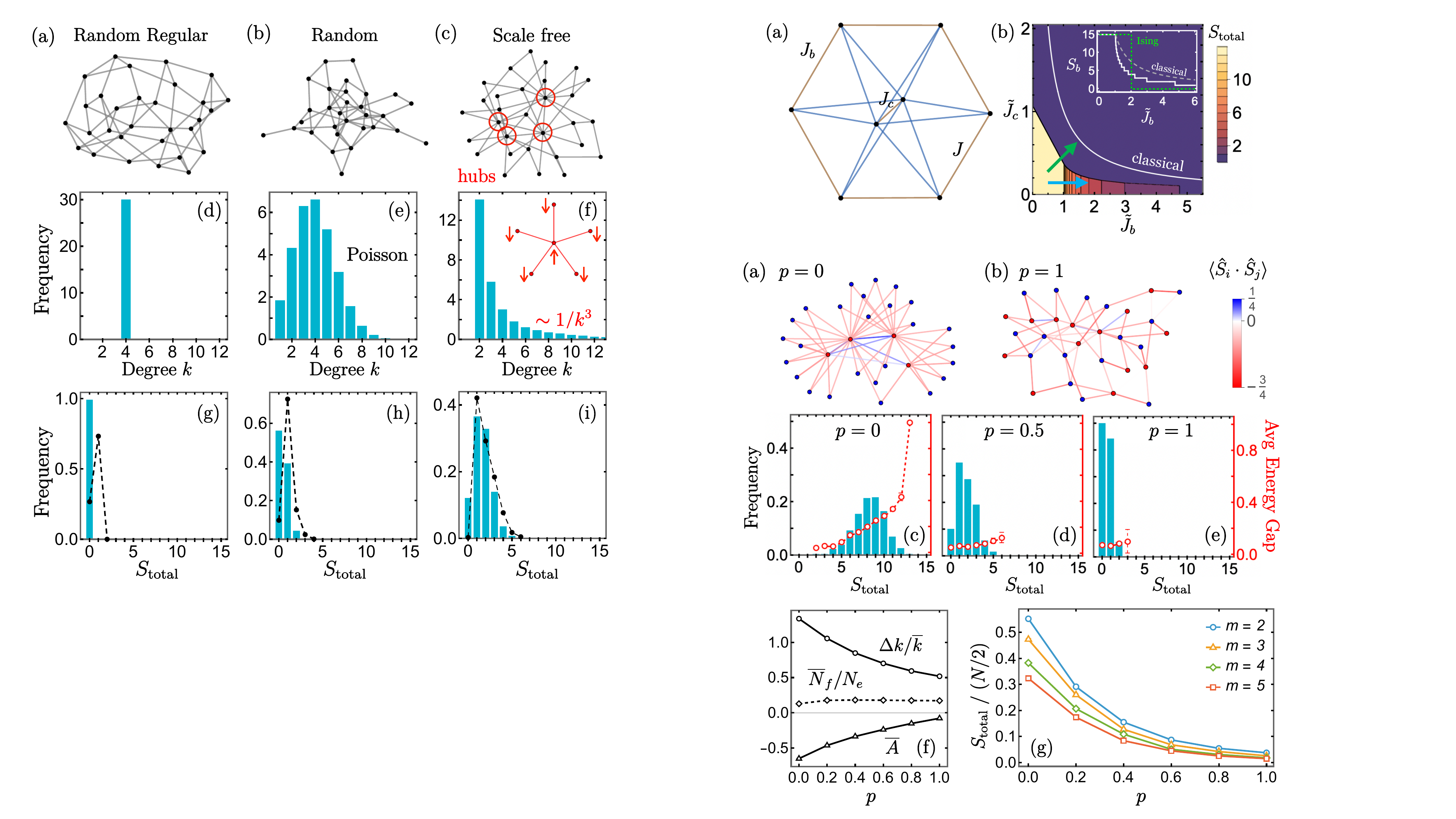}
    \caption{(a) Graph with embedded hubs and (b) graph with no hub from the ``copy model'' described in text for $m=2$. Node colors show a ground state for Ising spins and bond colors show ground-state correlations for our Heisenberg model. (c)--(e) Variation of total-spin distribution with the structure parameter $p$ for $N = 30$ and $m=2$. Red dots show the average excitation gap for a given $S_{\rm total}$. (f) Variation of heterogeneity $\Delta k$, frustration index $N_f$, and assortativity $A$. (g) $\bar{S}_{\rm total}$ as a function of $p$ and $\bar{k} = 2m - m(m+1)/N$ for $N = 30$.
    }
    \label{fig:copymodel}
\end{figure}

\textit{Tunable spin distribution}|We propose a simple model to grow networks with tunable heterogeneity and assortativity. It has a parameter $m \in \mathbb{Z}$ that sets the average degree and another parameter $p \in [0,1]$ that controls the structure. The protocol starts with a small clique of $m$ nodes, and in each step links a new node $i$ to $m$ existing nodes as follows: (1) Select a node $j$ at random, (2) with probability $p$ link $(i,j)$, (3) else link $i$ to a neighbor of $j$ with preferential attachment, i.e., with probability proportional to the neighbor's degree. This is a variant of the ``copy models'' for the Web \cite{Kleinberg1999, Kumar2000, Alam2019}.

As shown in Fig.~\ref{fig:copymodel}(a), for $p=0$ we get strongly disassortative graphs where many low-degree ``outer'' nodes connect to a few ``central'' hubs. As $p$ is increased to $1$, the graphs become more homogeneous [Fig.~\ref{fig:copymodel}(b)]. Thus, both heterogeneity and disassortativity fall off with $p$, while the frustration index is almost invariant [Fig.~\ref{fig:copymodel}(f)]. Accordingly, we find the total-spin distribution is peaked at a macroscopic value of $O(N)$ for $p=0$ and moves toward zero as $p$ goes to $1$ [Figs.~\ref{fig:copymodel}(c)-\ref{fig:copymodel}(e)]. In fact, finite-size scaling for $m=2$ indicates that this variation with $p$ spans the full range of $\bar{S}_{\rm total}$ for $N \to \infty$ (see End Matter). Such large, tunable magnetization in the presence of frustration is not found on regular lattices. %In addition, $\bar{S}_{\text{total}}$ decreases with the average degree $\bar{k} \approx 2m$ [Fig.~\ref{fig:copymodel}(g)] as in random graphs.

In Fig.~\ref{fig:copymodel}(c) the maximum total spin occurs when all of the outer nodes link to the initial clique, producing $m$ giant hubs that align opposite to the outer spins, yielding $S_{\text{total}} = N/2-m$. However, if a new node connects to an outer node, the latter can then accumulate more bonds to become a hub at the expense the initial clique, reducing $S_{\text{total}}$ [Fig.~\ref{fig:copymodel}(a)]. Note that the bonds between the hubs add frustration without changing the total spin.

In addition, we find that the excitation gap rises with $S_{\rm total}$, especially for $p=0$ [Figs.~\ref{fig:copymodel}(c)-\ref{fig:copymodel}(e)], showing that the hubs also lend more energetic stability. This may be anticipated as the gap for a ring is $O(1/N)$ \cite{spinwave1962}, whereas that for a star graph is $O(1)$. In most cases, we find the gap is magnetic, involving changes in $S_{\rm total}$ by $\pm 1$.

\begin{figure}
    \centering
    \includegraphics[width=1\columnwidth]{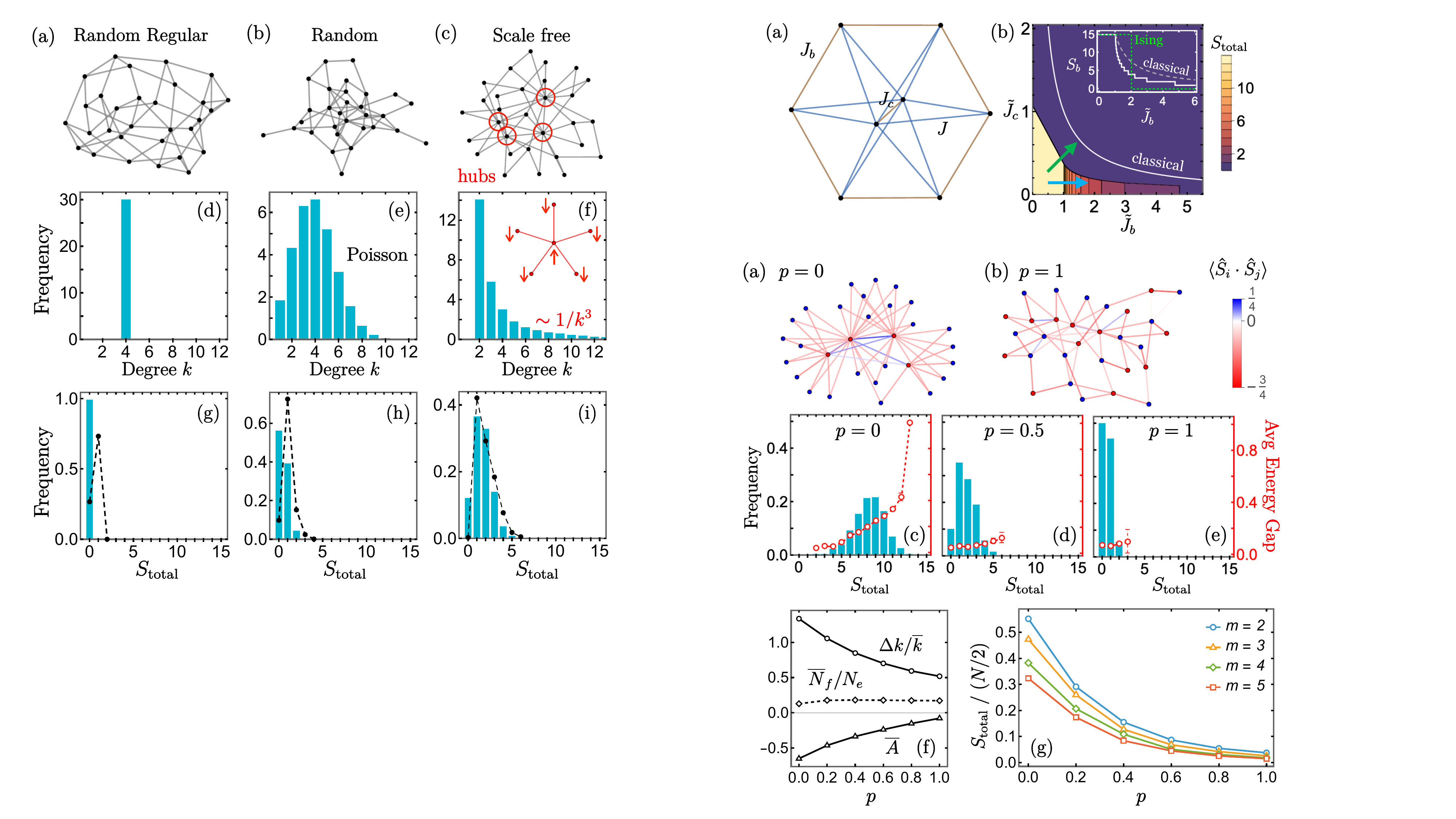}
    \caption{(a) Wheel with $N_c=2$ central spins and $N_b = 6$ outer spins. (b) Phase diagram for $N_b = 30$ where $\tilde{J}_b \coloneqq 4 J_b / (J N_c)$ and $\tilde{J}_c \coloneqq J_c N_c / (J N_b)$. Green arrow shows a discontinuous transition from a state where central and outer spins are antialigned to a state where both sum to zero. Blue arrow shows a continuous transition where the net outer spin $S_b$ falls in steps of $1$ at a rate $\sim N_b (\tilde{J}_b - 1)^{-1/2}$ (inset). This divergence is absent for Ising spins (dotted) and classical Heisenberg spins (dashed). For the latter, $S_{\text{total}} = 0$ for $\tilde{J}_b \tilde{J}_c > 1$ (solid curve).
    }
    \label{fig:wheel}
\end{figure}

\textit{Deterministic tuning of $S_{\text{total}}$ in a frustrated network}|Motivated by these results, we explore the phase diagram of a simple, hubbed network shaped like a wheel. It consists of $N_c$ central spins with all-to-all coupling $J_c$, $N_b$ outer spins forming a ring with coupling $J_b$, and all ``spokes'' connecting them with coupling $J$. Figure~\ref{fig:wheel}(a) shows a sketch for $N_c = 2$. One recovers distinct topologies as one of these couplings is reduced toward zero: For $J_b = 0$ and $J_c \sim J$, it is a $p=0$ graph [Fig.~\ref{fig:copymodel}(a)] with $S_{\rm total} = (N_b - N_c)/2 \gg 1$. For $J_c=0$, it is equivalent to a Heisenberg ring in a (quantized) magnetic field, which should undergo a transition from a saturated ferromagnet to a Luttinger liquid with decreasing total spin as $J_b/J$ is increased \cite{Kono2015}. For $J=0$, the ring is disconnected from the center, yielding $S_{\rm total} = 0$ (if $N_b$, $N_c$ are even). Thus, by varying the couplings one should be able to stabilize any total spin between $0$ and $(N_b - N_c)/2$.

These intuitions are confirmed by solving the system exactly using the Bethe Ansatz for a Heisenberg ring \cite{Karbach1998, Griffiths1964}. As shown in Fig.~\ref{fig:wheel}(b), the phase diagram features both abrupt and continuous transitions. Most notably, one finds a line of quantum critical points with power-law diverging susceptibility \cite{Kono2015, Karbach1998, Griffiths1964} at $J_b = J N_c / 4$, which is absent for classical Heisenberg spins. (See SM \cite{SuppMat} for derivations and more general comparisons between classical and quantum Heisenberg networks.) Note that Ising spins show only abrupt transitions from $S_{\rm total} = 0$ to $S_{\rm total} \geq (N_b - N_c)/2$  [Fig.~\ref{fig:wheel}(b) inset]. 
%Note that when either $N_b$ or $N_c$ is odd the ground state can be degenerate [beyond the usual ${\rm SU}(2)$ degeneracy] \cite{Karbach1995, Barwinkel2000, Curtright2017}.

\textit{Outlook}|Perhaps the most promising platform for realizing our setup is ion traps, for which there are concrete protocols for engineering arbitrary pairwise Heisenberg coupling with existing technology for several tens of qubits \cite{Korenblit2012, Teoh2020, Davoudi2020}. A large class of networks can also be fabricated in photonic platforms \cite{Tsomokos2010, Kollar2019}, where Heisenberg interactions may be simulated digitally \cite{Lamata2018, Hung2016} or using tailored light-matter coupling \cite{Kay2008}. 

Our findings have parallels in classical network science, where heterogeneity has been found to promote cooperative phenomena. For instance, it greatly increases the critical temperature for ferromagnetic Potts models and reduces the synchronization threshold for Kuramoto oscillators \cite{Dynamics_book, Kuramoto_2016}. These correlate with a large ground-state total spin, which provides a natural way to synchronize by adding damping and a magnetic field that rotates the spins. Furthermore, we show that the impact of heterogeneity can be hugely enhanced by adding disassortative correlations (Fig.~\ref{fig:assortativity}), and that quantum effects can yield a qualitatively different response (Fig.~\ref{fig:wheel}). These results strongly encourage the exploration of collective dynamical phenomena on quantum many-body networks.

They also raise a number of open questions on frustrated magnetism: (1) Why is the total spin generally insensitive to frustration level when all bonds are equal? This is far from obvious and does not hold for kinetic magnetism \cite{Revathy_2025}, where magnetic order arises from the path interference of a hole or doublon \cite{Nagaoka_review}. (2) What is the nature of the ground state? Ising spins exhibit glassy behavior on small-world and scale-free graphs \cite{2008_RMP, SG_2008, SG_2006, SG_2009}. Can quantum fluctuations in the Heisenberg model stabilize a spin liquid in the absence of lattice symmetry \cite{Kimchi2018, Balents2010}? Answering this question would require a more detailed examination of the spin correlations. While we generally find antiferromagnetic two-site correlations, the correlation length grows with disassortativity, and second or third neighbors can be more strongly correlated than adjacent spins on hubbed networks (see End Matter). (3) How are the dynamical and ergodicity properties affected by network topology \cite{SW_Scrambling_2019, Chaos_Ising_Adolfo_2025}? (4) Is the magnetic order sensitive to community structures or other motifs \cite{Motifs} that lead to hierarchical or explosive synchronization in classical networks \cite{Dynamics_book, Kuramoto_2016, Explosive_2019, Explosive_2016, Cluster_2021}?

%We hope our findings herald such broad explorations at the intersection of complex networks and quantum many-body physics.

\vspace{1em}
\begin{acknowledgments}
{\it Acknowledgments}|We thank Sitabhra Sinha, Sthitadhi Roy, Claudio Castelnovo, and Subhro Bhattacharjee for useful discussions and comments. We thank the National Supercomputing Mission (NSM) for providing computing resources of ``PARAM Utkarsh'' at CDAC Knowledge Park Bangalore, which is implemented by C-DAC and supported by the Ministry of Electronics and Information Technology (MeitY) and Department of Science and Technology (DST), Government of India.

\vspace{1em}
{\it Data availability}|The data and source codes that support our findings are openly available \cite{dataset, source_codes}.
\end{acknowledgments}

% \begingroup
% \renewcommand{\addcontentsline}[3]{}% Remove functionality of \addcontentsline
% \renewcommand{\section}[2]{}% Remove functionality of \section


%\endgroup

% Force balance columns for bibliography before end matter
\twocolumngrid

% Start end matter
\onecolumngrid
\vspace{3em}
\begin{center}
    \textbf{End Matter}
\end{center}
\vspace{1em}
\twocolumngrid

\appendix

\begin{figure}[!htb]
    \centering
    \includegraphics[width=1\columnwidth]{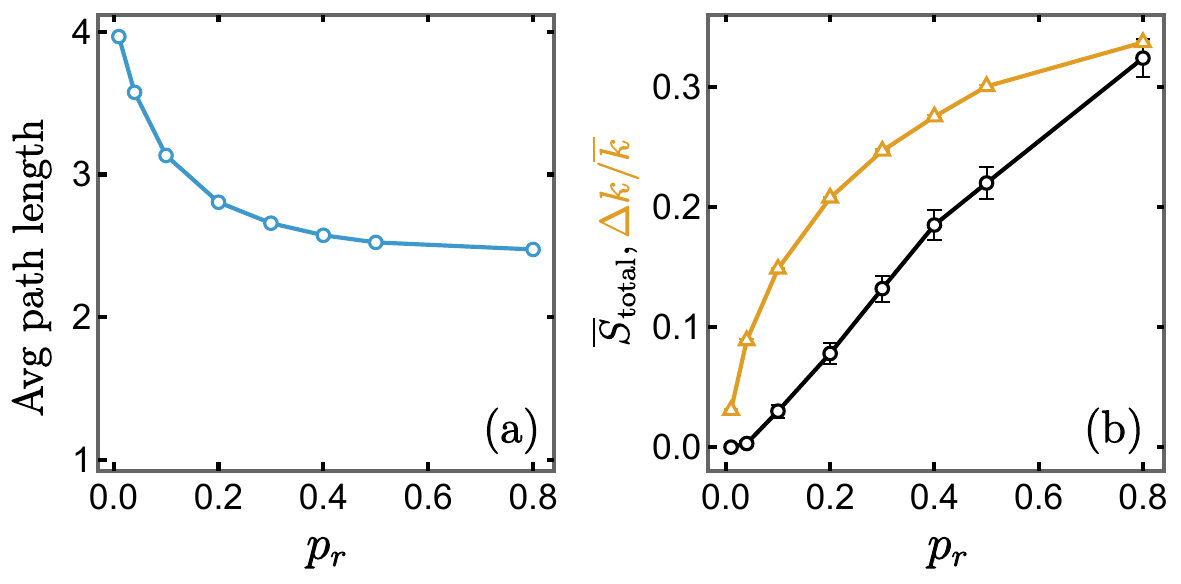}
    \caption{(a) Path length, (b) total spin and heterogeneity as a function of the rewiring probability $p_r$ for Watts-Strogatz small-world graphs with $N = 30$ and $\bar{k} = 4$.}
    \label{endfig:smallworld}
\end{figure}

{\it Small-world networks}|One can interpolate between a regular lattice and random graphs by rewiring a small fraction of the bonds \cite{WattsStrogatz1998}; These ``shortcuts'' dramatically shorten the average path length between two nodes while maintaining a high clustering, resulting in a small-world character. Figure~\ref{endfig:smallworld} shows that this procedure leads to a slow increase of the average total spin, accompanied by a similar rise in the heterogeneity. Thus, we find no strong dependence of $S_{\text{total}}$ on the path length. This is similar to the critical temperature of a ferromagnetic Ising model on small-world networks, which increases from zero to a finite value as more bonds are rewired \cite{Barrat2000}.

\begin{figure}[!htb]
    \centering
    \includegraphics[width=0.98\columnwidth]{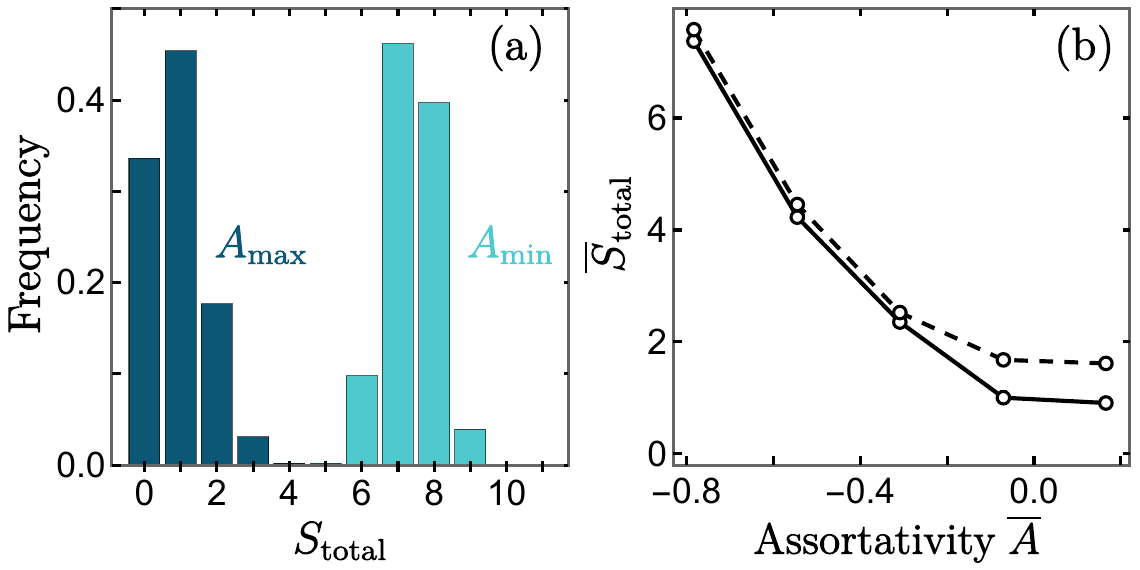}
    \caption{(a) Total-spin distribution of the most assortative ($\bar{A}=+0.17$) and the most disassortative ($\bar{A}=-0.78$) scale-free graphs with $N = 30$ and $\bar{k} = 3.8$. (b) Variation of $\bar{S}_{\text{total}}$ with $\bar{A}$. Dashed curve shows $\bar{S}^z_{\rm total}$ for Ising spins.
    }
    \label{endfig:scalefree}
\end{figure}

\vspace{1em}
{\it (Dis)assortative scale-free networks}|In Fig.~\ref{fig:assortativity} we explained how increasing the assortativity coefficient of random graphs causes $S_{\text{total}}$ to fall sharply. Figure~\ref{endfig:scalefree} shows the same is true for Barab{\'a}si-Albert graphs with scale-free degree distribution ($p_k \sim 1/k^3$).

\begin{figure}[!b]
    \centering
    \includegraphics[width=0.98\columnwidth]{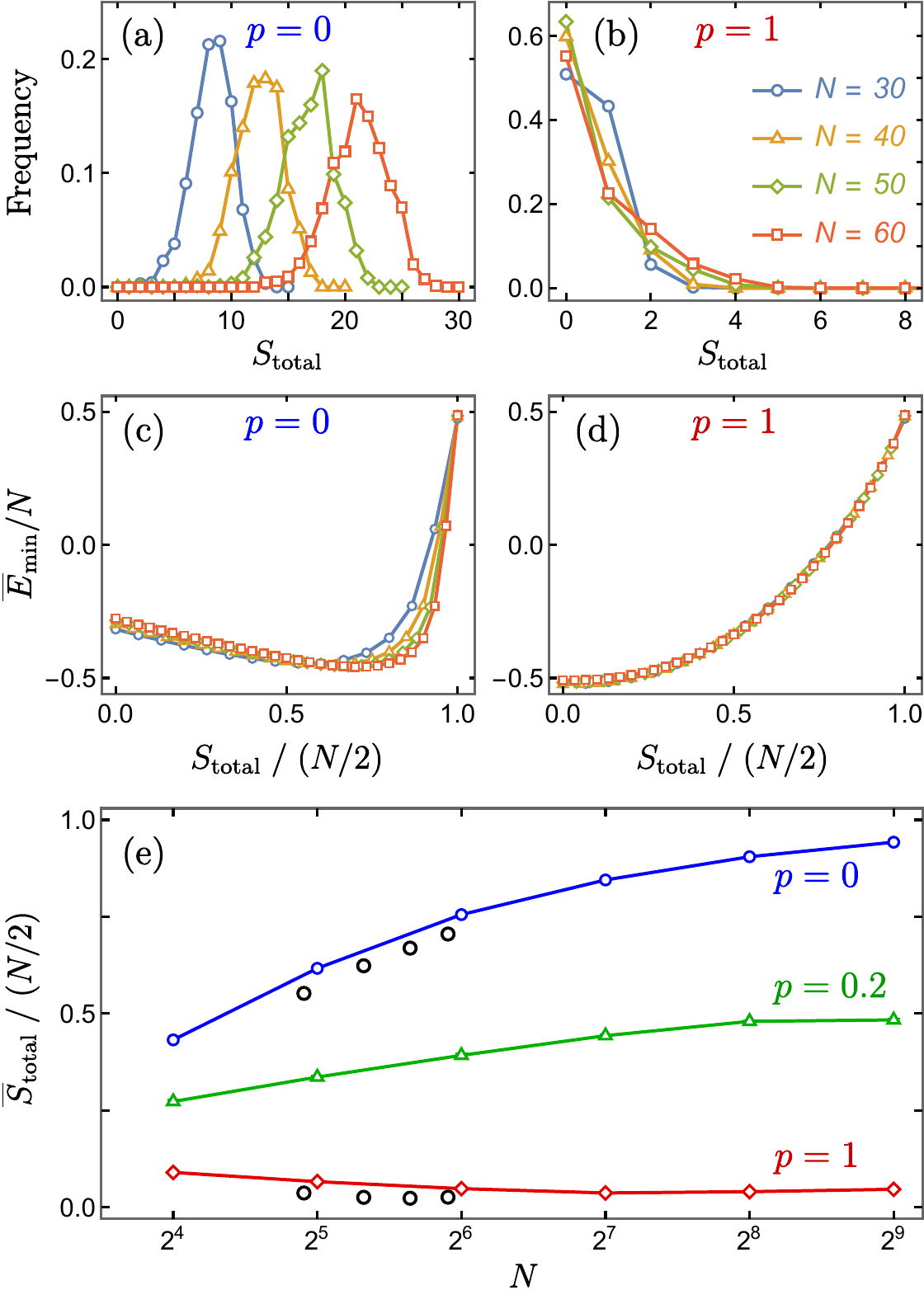}
    \caption{(a,b) Total-spin distribution with increasing system size $N$ for the copy model with $m=2$. (c,d) Minimum energy as a function of $S_{\rm total}$, averaged over the corresponding ensembles. (e) Average total spin of the ground state. Black circles show $\bar{S}_{\rm total}$ for the quantum Heisenberg model and solid lines show $\bar{S}^z_{\rm total}$ for Ising spins (max-cut bipartitions).
    }
    \label{endfig:FSS}
\end{figure}

\vspace{1em}
{\it Finite-size scaling}|In Fig.~\ref{fig:copymodel} we showed how the spin distribution is tunable over a macroscopic range by varying the parameter $p$ of the copy model, which controls heterogeneity and assortativity levels. To check how this result depends on system size $N$, we have performed finite-size scaling for $m=2$ using total-spin resolved matrix product states in \texttt{Block2} \cite{block2}, which agrees with exact diagonalization for $N \leq 30$. For the strongly hubbed networks ($p=0$), as $N$ is increased up to $60$, the distribution shifts toward the maximum total spin [Fig.~\ref{endfig:FSS}(a)] and the energy vs $S_{\rm total}$ profile becomes sharper [Fig.~\ref{endfig:FSS}(c)]. In contrast, for $p=1$ (no hub, low total spin), the states are more entangled and the simulations did not always converge within a bond dimension of $1024$ \cite{DMRG_2011}. However, up to this bound, the distribution remains concentrated near $S_{\rm total}=0$ [Figs.~\ref{endfig:FSS}(b) and \ref{endfig:FSS}(d)]. These trends are also seen for $S^z_{\rm total}$ of Ising spins (max-cut bipartitions) up to $N=512$ [Fig.~\ref{endfig:FSS}(e)]. So we infer that one can access the full range of $S_{\rm total}$ by varying $p$ for large $N$. In other words, the central role of heterogeneity and assortativity is magnified by increasing system size.

The average $S^z_{\rm total}$ for Ising spins was computed from up to $1000$ degenerate solutions for each graph using the Gurobi Optimizer \cite{Aref2019, gurobi}.

\begin{figure}[b]
    \centering
    \includegraphics[width=1\columnwidth]{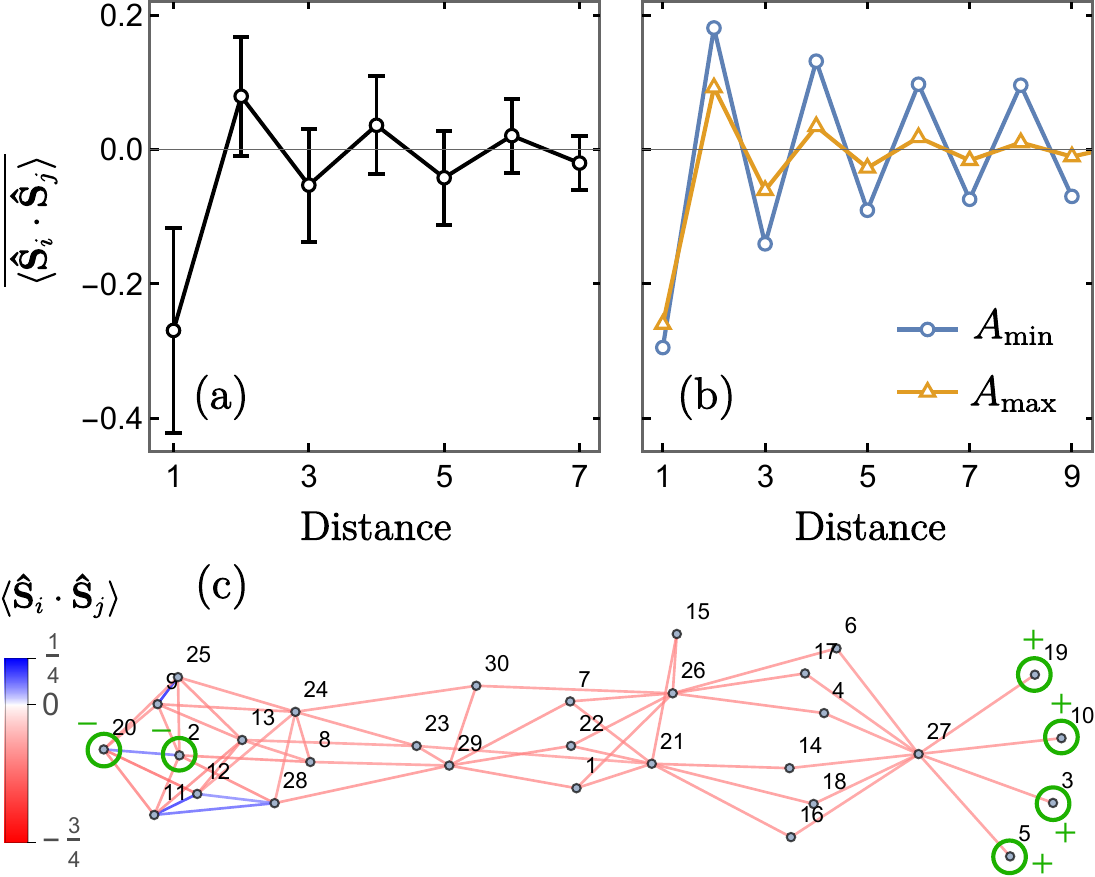}
    \caption{(a) Spin correlation as a function of graph distance for random graphs with $N=30$ and $\bar{k} = 4$. Dots show the average values and error bars show the standard deviations. (b) Effect of tuning the assortativity to its minimum ($\bar{A} = -0.9$) and maximum ($\bar{A} = +0.7$) values. A similar variation is found for the scale-free Barab{\'a}si-Albert networks. (c) A disassortative random graph ($A = -0.95$) with negative correlations of $-0.13 \pm 0.02$ between sites $\{2,20\}$ and $\{3,5,10,19\}$, circled with opposite signs, which are $7$th neighbors of each other.
    }
    \label{endfig:correlation}
\end{figure}

\begin{figure*}[!htb]
    \centering
    \includegraphics[width=\textwidth]{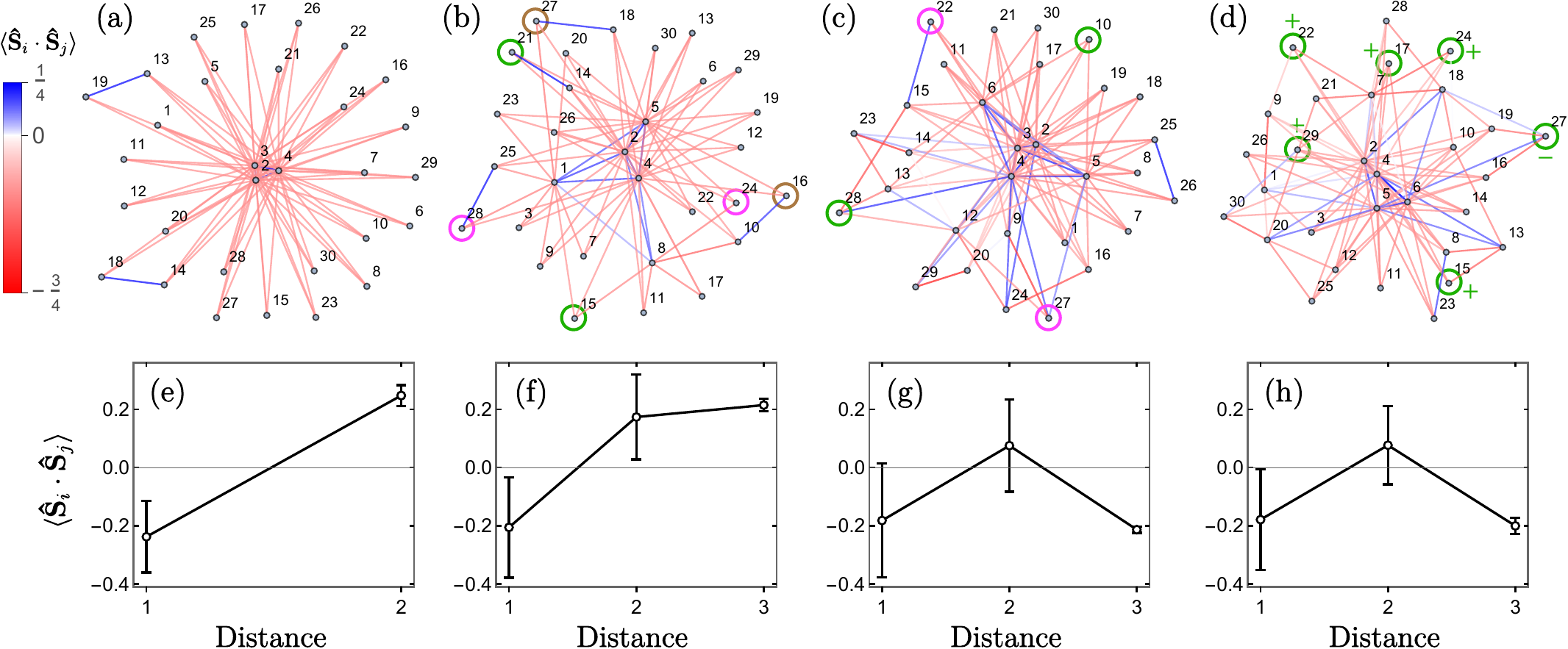}
    \caption{(a-d) Copy-model networks with three or more hubs where the average correlation between second or third neighbors is stronger than that between nearest neighbors. (e-f) The corresponding spin correlations as a function of graph distance, showing the averages (dots) and the standard deviations (error bars). In (a), all the periphery nodes are strongly aligned. In (b) and (c), the nodes circled with the same color are third neighbors of each other with strong positive and negative correlations, respectively. In (d), node $27$ is strongly anti-aligned with its third neighbors: nodes $15$, $17$, $22$, $24$, and $29$.
    }
    \label{endfig:correlation_examples}
\end{figure*}

\vspace{1em}
{\it Long-range correlations}|While we have focused on the total spin, we have also analyzed the two-site spin correlations as a function of their graph distance (path length). We generally find antiferromagnetic correlations that are strongest between nearest neighbors; however, the correlation length is markedly increased by making the connectivity more disassortative, as shown in Fig.~\ref{endfig:correlation}. This is because the hubs can sustain a relatively strong correlation between the ``high-low'' and ``mid-mid'' regions [Figs.~\ref{endfig:correlation}(c) and \ref{fig:assortativity}]. In addition, for the copy-model networks with multiple hubs ($m \gtrsim 3$, small $p$), we find several cases where the correlation between two periphery sites, which are second or third neighbors of each other, is stronger than that between adjacent sites. Moreover, this correlation can be positive or negative depending on whether their neighbors are aligned (e.g., they connect to the same hubs) or anti-aligned (they connect to different hubs or other periphery sites). Figure~\ref{endfig:correlation_examples} shows a few representative examples.

\onecolumngrid
\clearpage
\renewcommand{\baselinestretch}{1.3}\normalsize
\begin{center}
\textbf{\large Supplemental Material for\\
	    ``Frustrated Quantum Magnetism on Complex Networks: What Sets the Total Spin''}\\
\vspace{0.5cm}
Preethi Gopalakrishnan and Shovan Dutta
\end{center}
\vspace{1cm}

\twocolumngrid

%%%%%%%%%%%%%%%%%%%%%%%%%%%%%%%%%%%%%%%%%%%%%%%%%%%%%

% Arrange section headings, equation numbers, figure numbers etc to start with S, e.g., Figure S1, S2, etc.  

    \setcounter{table}{0}
    \renewcommand{\thetable}{S\arabic{table}}%
    \setcounter{figure}{0}
    \renewcommand{\thefigure}{S\arabic{figure}}%
    \renewcommand{\theHfigure}{S\arabic{figure}}
    \setcounter{secnumdepth}{3}
    \setcounter{section}{0}
    \renewcommand{\thesection}{\Roman{section}}%
    \setcounter{subsection}{0}
    \renewcommand{\thesubsection}{\Alph{subsection}}
    \setcounter{equation}{0}
    \renewcommand{\theequation}{S\arabic{equation}}%
    \setcounter{page}{1}
    \renewcommand{\thepage}{SM-\arabic{page}}
    \renewcommand{\bibnumfmt}[1]{[S#1]}
    \renewcommand{\citenumfont}[1]{S#1}

%%%%%%%%%%%%%%%%%%%%%%%%%%%%%%%%%%%%%%%%%%%%%%%%%%%%%

\makeatletter
\renewcommand{\baselinestretch}{1}\normalsize
%\tableofcontents

\section{Random trees}
\begin{figure}[!htb]
\centering
    \includegraphics[width=1\columnwidth]{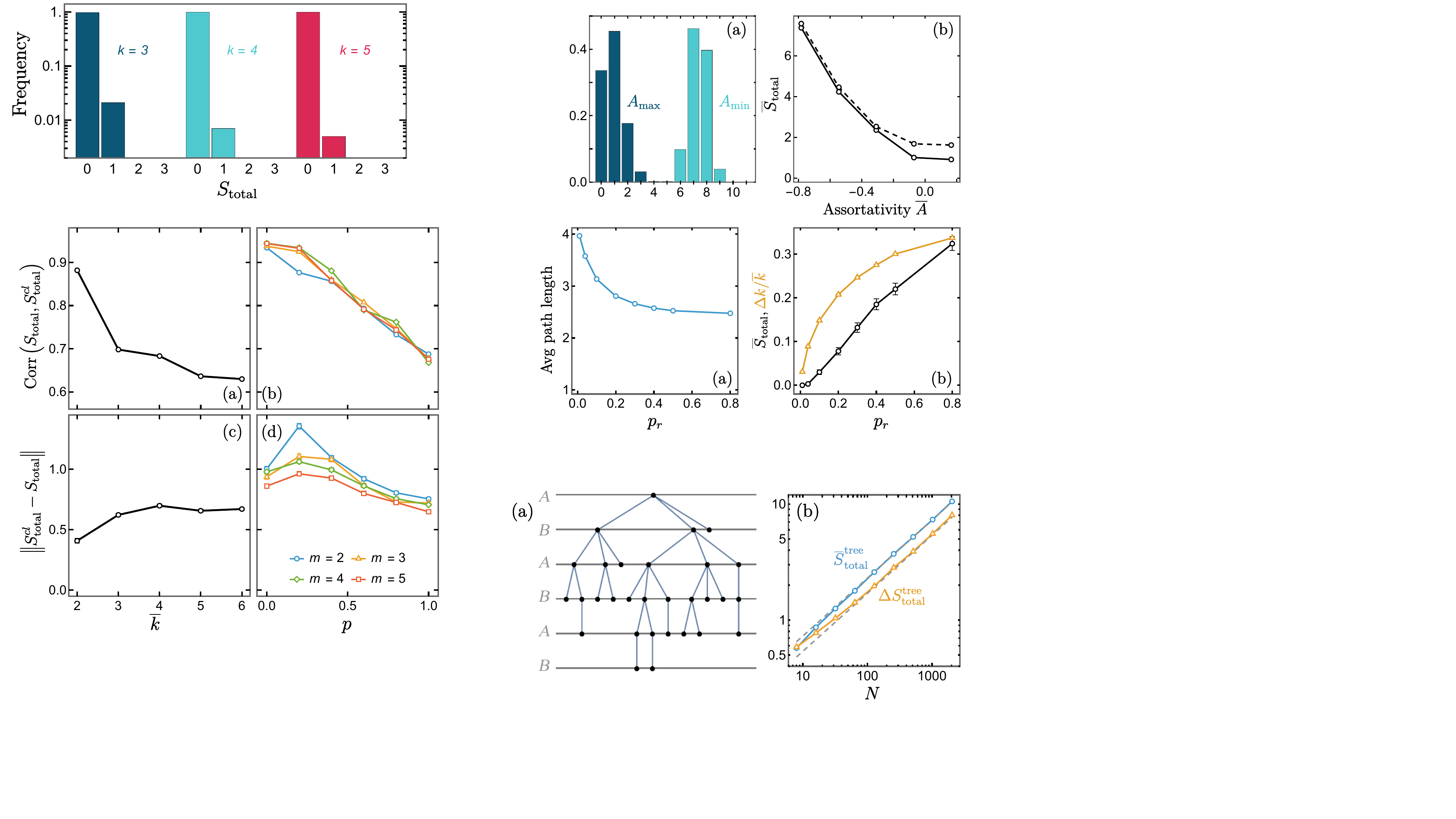}
    \caption{(a) Tree with $N=30$ sites. Horizontal lines show the two sublattices. (b) The average and the standard deviation of $S_{\text{total}}$ over ensembles of $5000$ random trees. Dashed lines show the fits $0.23\sqrt{N}$ and $0.17\sqrt{N}$, respectively.
    }
    \label{suppfig:tree}
\end{figure}

A tree has a bipartite structure [Fig.~\ref{suppfig:tree}(a)], for which $S_{\text{total}} = |N_A - N_B| / 2$, where $N_A$ and $N_B$ are the numbers of spins in the two sublattices \cite{supp_Lieb1962}. Figure~\ref{suppfig:tree}(b) shows that both the average and the spread of $S_{\text{total}}$ over random trees with $N$ sites and $N-1$ bonds  scale as $\sqrt{N}$.

\section{Random regular graphs}
\begin{figure}[!htb]
    \centering
    \includegraphics[width=1\columnwidth]{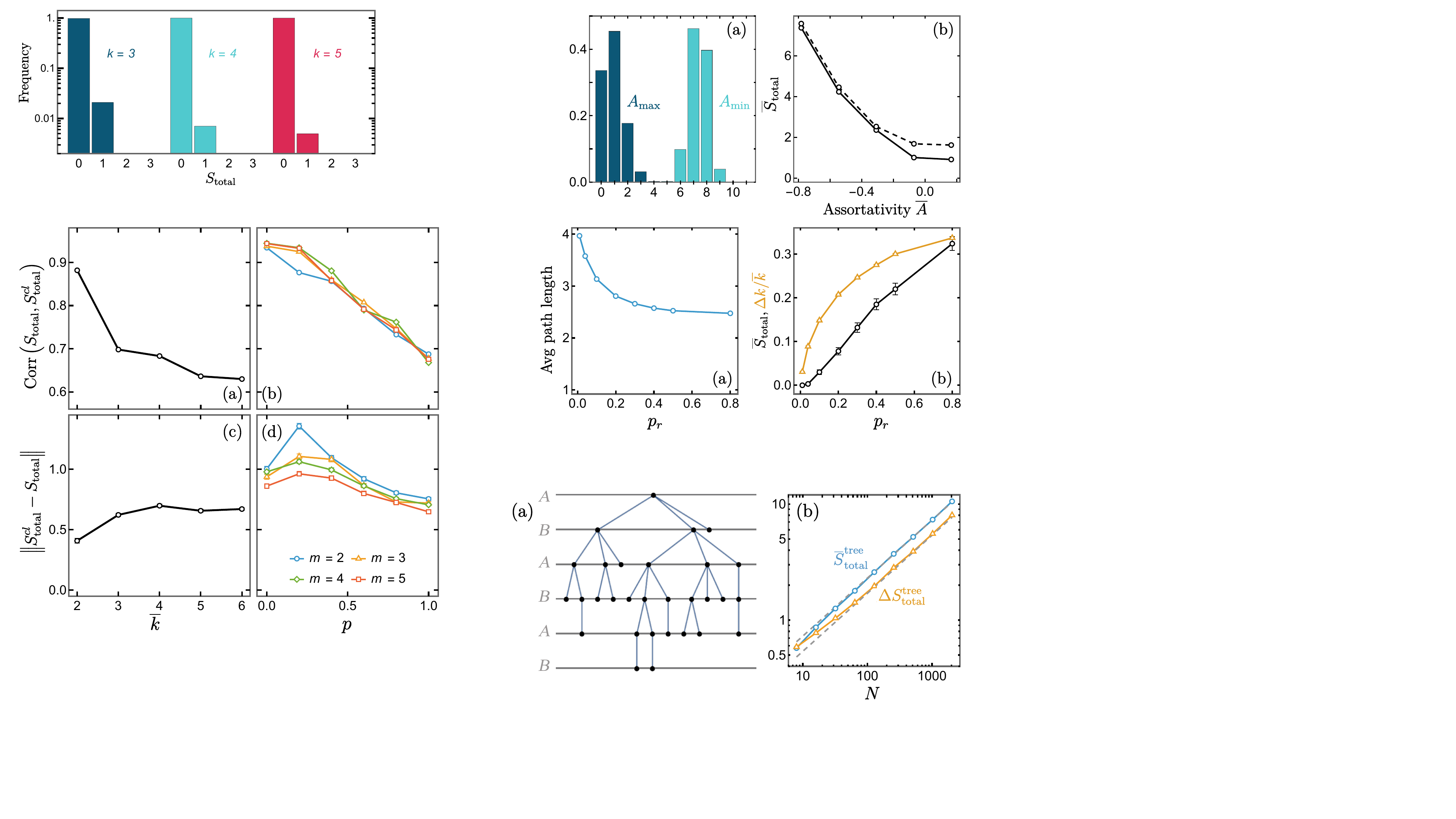}
    \caption{Total-spin distribution of random regular networks with $N=30$ sites for different values of the degree $k$.
    }
    \label{suppfig:randomregular}
\end{figure}

In the main text we showed that a random regular graph, where each node has $k=4$ neighbors, almost always has $S_{\text{total}} = 0$ in ground state. Figure~\ref{suppfig:randomregular} shows the same holds for $k=3$ and $k=5$, further demonstrating that heterogeneity is essential for nonzero total spin.

\section{Quantum vs.~classical $S_{\text{total}}$}
\begin{figure}[!htb]
    \centering
    \includegraphics[width=1\columnwidth]{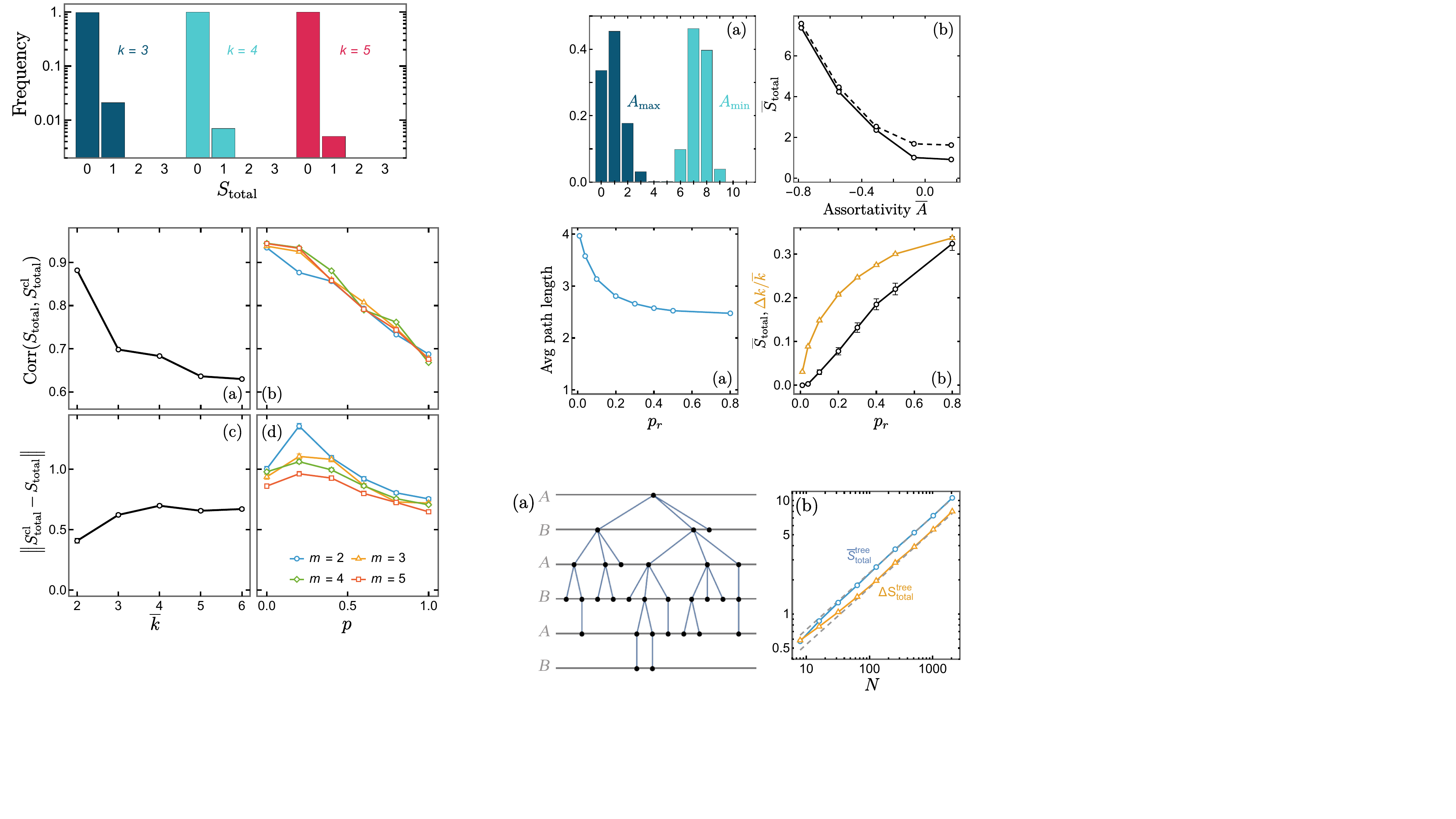}
    \caption{(Top panel) Correlation and (bottom panel) root-mean-square deviation between the net magnetizations of classical and quantum spins for random graphs (left column) and copy-model graphs (right column) with $N = 30$ sites.
    }
    \label{suppfig:classicalVquantum}
\end{figure}

Figure~\ref{suppfig:classicalVquantum} shows a comparison between the total spins of classical and quantum Heisenberg models defined on two types of networks: (i) random graphs and (ii) graphs generated by the copy model introduced in the main text, where heterogeneity and assortativity can be tuned by varying a structure parameter $p$. Generally, we find that the two spins are highly correlated and that the classical total spin is larger by $O(1)$.

\begin{figure*}[!htb]
    \centering
    \includegraphics[width=\textwidth]{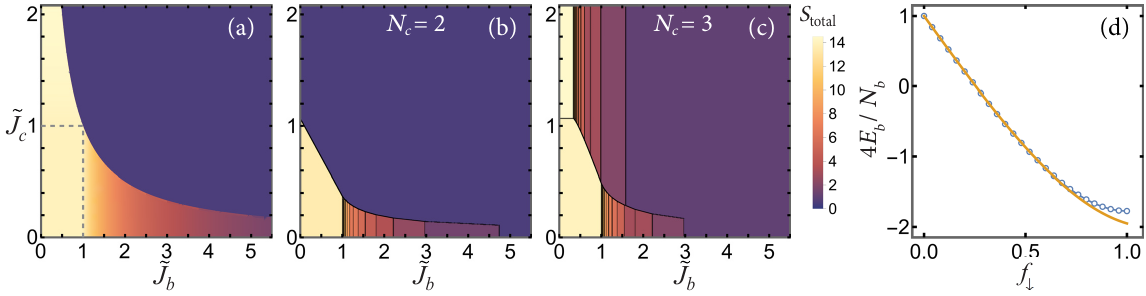}
    \caption{Phase diagram of the wheel-shaped network introduced in the main text [Eq.~\eqref{suppeq:wheelhamil}] for (a) classical spins with $N_c = 2$, $N_b = 30$, (b) qubits with $N_c = 2$, $N_b = 30$, and (c) qubits with $N_c = 3$, $N_b = 30$. (d) Minimum energy of the Heisenberg ring as a function of twice the fraction of $\downarrow$ spins for $N_b = 50$. Orange curve shows the fit $1 - 4 f_{\downarrow} + 1.05 f_{\downarrow}^3$.
    }
    \label{suppfig:wheel}
\end{figure*}

\section{Exact solution for the wheel graph}
In the main text we proposed a wheel-shaped network where the total spin can be tuned by varying some of the bond strengths. The network consists of $N_c$ central spins and $N_b$ outer spins with the Hamiltonian
\begin{equation}
    \hat{H} = \frac{J_c}{2} \hat{\mathbf{S}}_c^2 
    + J \hat{\mathbf{S}}_b \cdot \hat{\mathbf{S}}_c 
    + J_b \sum_{n=1}^{N_b} \hat{\mathbf{S}}_n \cdot \hat{\mathbf{S}}_{n+1} 
    \label{suppeq:wheelhamil}
\end{equation}
up to a constant, where $\hat{\mathbf{S}}_c$ is the net central spin, $\hat{\mathbf{S}}_b \coloneqq \sum_{i=1}^{N_b} \hat{\mathbf{S}}_i$ is the net outer spin, and $\hat{\mathbf{S}}_{N_b+1} \equiv \hat{\mathbf{S}}_1$. Below we describe exact solutions for the ground states of this network for both classical and quantum spins.

\subsection{Classical spins}
In order to lower energy $\mathbf{S}_b$ and $\mathbf{S}_c$ point in opposite directions. Without loss of generality we take $\mathbf{S}_b$ along $+z$ and $\mathbf{S}_c$ along $-z$. The maximum magnitude of $\mathbf{S}_c$ is $N_c S$, where $S$ is the magnitude of each spin. To compare with qubits we set $S=1/2$ and write $\mathbf{S}_c = -(N_c/2) u \hat{\mathbf{z}}$, where $u \in [0,1]$. This central spin acts as an effective magnetic field for the Heisenberg ring [see Eq.~\eqref{suppeq:wheelhamil}], whose thermodynamic properties can be solved exactly \cite{supp_Blume1965}. The field wants to polarize the spins along $+z$, whereas the bonds prefer a N\'{e}el-type antiferromagnet. In the ground state, the spins orient along the polar and azimuthal angles
\begin{subequations}
\begin{align}
    \theta_n &= \theta \;,\\
    \phi_n &= n (\pi - \alpha) + \phi_0 \;,
\end{align}
\end{subequations}
where $\theta$ goes from $0$ to $\pi/2$ as the field decreases, $\phi_0$ is an arbitrary constant, and $\alpha = \big[1-(-1)^{N_b}\big] \pi / (2 N_b)$. This canted antiferromagnet has energy
\begin{equation}
    \Tilde{E} = \Tilde{J}_c u^2 - 2 u v + \Tilde{J}_b \cos^2(\alpha/2) v^2 \;,
\end{equation}
where $v \coloneqq \cos \theta$, $\Tilde{E} \coloneqq 8 E/(J N_b N_c)$, $\Tilde{J}_c \coloneqq J_c N_c/(J N_b)$, and $\Tilde{J}_b \coloneqq 4 J_b/(J N_c)$. Minimizing $\Tilde{E}$ with respect to the fractions $u$ and $v$ yields a ground state with $S_b = \frac{1}{2}N_b v$, $S_c = \frac{1}{2}N_c u$, and $S_{\text{total}} = \frac{1}{2}|N_b v - N_c u|$.

The resulting phase diagram comprises four regions:
\begin{subequations}
\begin{align}
    u = v = 0 \quad &\text{if } 
    \Tilde{J}_c \Tilde{J}_b^{\prime} > 1 \;, \\
    u = v = 1 \quad &\text{if } 
    \Tilde{J}_c < 1 \text{ and } \Tilde{J}_b^{\prime} < 1 \;, \\
    u = 1, v = 1/\Tilde{J}_b^{\prime} \quad &\text{if }
    \Tilde{J}_c \Tilde{J}_b^{\prime} < 1 \text{ and } \Tilde{J}_b^{\prime} > 1 \;, \\
    v = 1, u = 1/\Tilde{J}_c \quad &\text{if }
    \Tilde{J}_c \Tilde{J}_b^{\prime} < 1 \text{ and } \Tilde{J}_c > 1 \;,
\end{align}
\end{subequations}
with $\Tilde{J}_b^{\prime} \coloneqq \Tilde{J}_b \cos^2(\alpha/2)$, as shown in Fig.~\ref{suppfig:wheel}(a) for $N_b \gg N_c$. Note that $\alpha \approx 0$ for large $N_b$ and the phase diagram does not depend on whether $N_c$ is odd or even.

\subsection{Quantum spins}
Since $\hat{\mathbf{S}}_b^2$ and $\hat{\mathbf{S}}_c^2$ commute with $\hat{H}$ and $S_{\text{total}} = S_{bc} \coloneqq |S_b - S_c|$ in the ground state(s), we can write the lowest energy for a given $S_b$ and $S_c$ as
\begin{equation}
    E = \frac{J_c}{2} \hat{S}_c^2 
        + \frac{J}{2} \big(\hat{S}_{bc}^2 - \hat{S}_b^2 - \hat{S}_c^2 \big)
        + J_b E_b (N_b, S_b) \;,
    \label{suppeq:Ewheel}
\end{equation}
where $\hat{S}^2 \coloneqq S(S+1)$ and $E_b$ is the lowest energy of the Heisenberg ring. The latter can be found using the Bethe Ansatz \cite{supp_Karbach1998}, which involves solving the equations
\begin{equation}
    N_b \tan^{-1}(2 \lambda_i) = \pi I_i + \sum_j \tan^{-1} (\lambda_i - \lambda_j)
\end{equation}
for the rapidities $\lambda_i$, $i=1, 2, \dots, N_{\downarrow} = N_b/2 - S_b$, where $I_i = i - (N_{\downarrow}+1)/2$ for even $N_b$ and $i - (N_{\downarrow}+1 \pm 1)/2$ for odd $N_b$ \cite{supp_Karbach1995}, whereupon $E_b = N_b/4 - \sum_i 2/(4 \lambda_i^2 + 1)$. The two sets of $\{I_i\}$ for odd $N_b$ are degenerate in energy \cite{supp_Barwinkel2000}. The ground-state spins are obtained by minimizing $E$ in Eq.~\eqref{suppeq:Ewheel} over all quantum numbers $S_b = N_b/2, N_b/2-1, \dots$ and $S_c = N_c/2, N_c/2-1, \dots$.

Figures~\ref{suppfig:wheel}(b) and \ref{suppfig:wheel}(c) show that the resulting phase diagram differs qualitatively for $N_c = 2$ and $N_c = 3$. This is because for odd $N_c$ one cannot have $S_c = 0$, which leads to a stepwise fall of $S_b$ as $J_b$ is increased relative to $J$. To understand this feature more precisely we consider the energy as a function of $S_b$ for a given $S_c < S_b$,
\begin{equation}
    \Tilde{E} = \text{constant} + \frac{4S_c}{N_c} f_{\downarrow} + \frac{\Tilde{J}_b}{2} \Tilde{E}_b (f_{\downarrow}) \;,
\end{equation}
where $f_{\downarrow} \coloneqq 1 - 2 S_b / N_b$ is twice the fraction of $\downarrow$ spins in the ring, and $\Tilde{E}_b \coloneqq 4 E_b / N_b$. For large $S_b$ (small $f_{\downarrow}$) one can write $\Tilde{E}_b \approx 1 - 4 f_{\downarrow} + \eta f_{\downarrow}^3$, where $\eta \sim O(1)$ [Fig.~\ref{suppfig:wheel}(d)], which predicts the minimum energy for 
\begin{equation}
    f_{\downarrow} \approx \big[ 4 \big( \Tilde{J}_b - \Tilde{J}_b^* \big) / \big( 3\eta\Tilde{J}_b \big) \big]^{1/2}
\end{equation}
when $\Tilde{J}_b > \Tilde{J}_b^* \coloneqq 2S_c / N_c$. Consequently, the susceptibility $\partial S_b / \partial \Tilde{J}_b$ diverges as
\begin{equation}
    \frac{\partial S_b}{\partial \Tilde{J}_b} \approx
    - \frac{N_b}{2 \sqrt{3\eta} \Tilde{J}_b^*} 
    \bigg(\frac{\Tilde{J}_b}{\Tilde{J}_b^*} - 1\bigg)^{\! -1/2},
\end{equation}
which is seen in Fig.~\ref{suppfig:wheel}(b) for $S_c = 1$ and in Fig.~\ref{suppfig:wheel}(c) for $S_c = 3/2, 1/2$. No such divergence is present for classical spins [Fig.~\ref{suppfig:wheel}(a)]. This disparity also arises for a Heisenberg ring in a classical magnetic field \cite{supp_Griffiths1964, supp_Karbach1998, supp_Kono2015, supp_Blume1965}.

\end{document}